\documentclass{aastex}          
\usepackage{spr-astr-addons}    

\usepackage{amsmath} 
\newcommand{\vect}[1]{\boldsymbol{#1}} 
\usepackage{mathrsfs}
\usepackage{natbib}

\begin{document}

\title{Origin of the Cosmological Constant}

\shorttitle{Origin of the Cosmological Constant}
\shortauthors{J.O. Stenflo>}

\author{J.O. Stenflo\altaffilmark{1,2}}
\email{stenflo@astro.phys.ethz.ch} 

\altaffiltext{1}{Institute for Particle Physics and Astrophysics, ETH
  Zurich, CH-8093 Zurich, Switzerland}
\email{stenflo@astro.phys.ethz.ch}
\altaffiltext{2}{Istituto Ricerche Solari Locarno (IRSOL), Via Patocchi, CH-6605 Locarno-Monti, Switzerland}

\begin{abstract}
The observed value of the cosmological constant corresponds to a time
scale that is very close to the current conformal age
of the universe. Here we show that this is not a 
coincidence but is caused by  
a periodic boundary condition, which only manifests itself when the
metric is represented in Euclidian spacetime. The circular property of
the metric in Euclidian spacetime becomes an exponential evolution (de
Sitter or $\Lambda$ term) in ordinary spacetime. The value of
$\Lambda$ then gets uniquely linked to the period in Euclidian
conformal time, which corresponds to the conformal age of the universe. Without
the use of any free model parameters we predict 
the value of the dimensionless parameter $\Omega_\Lambda$ to be 67.2\,\%,
which is within $2\sigma$ of the value derived from CMB
observations. 

\end{abstract}


\section{Introduction}\label{sec:intro}
The cosmological constant was introduced by \citet{stenflo-einstein17}
to allow for the possibility of a static universe but was dismissed as
a mistake after the expansion of the universe was
discovered. \citet{stenflo-weinberg1987} showed that the magnitude of
the cosmological constant must be very small to permit our existence as
observers. This anthropic argument led to speculations about the
existence of parallel universes with a variety of values of the
cosmological constant. The accelerated expansion of
the universe that was discovered through the use of supernovae type Ia as standard candles
\citep{stenflo-riessetal1998a,stenflo-perlmutteretal1999a} could only be
modeled in terms of Friedmann-Lema\^{i}tre cosmology by
using the cosmological constant as a free parameter chosen to fit the
observations. 

While the cosmological constant $\Lambda$ is very useful as a model fitting
parameter, its physical nature has remained enigmatic. If it is
interpreted as a physical field with an energy density $\rho_\Lambda$,
then the circumstance that $\rho_\Lambda$ does not seem to depend on
redshift or look-back time leads to a major conceptual problem. If its
value remains constant with the evolution of the universe, then it
would be an extraordinary coincidence that its value happens to be of
the same order of magnitude as the current matter density $\rho_M$. In
the past its role would have been insignificant, while in the future
it would completely dominate and drive an exponential expansion of the
universe. This would violate the Copernican principle, which says that
we are not privileged observers. 

The aim of the present paper is to show that the accelerated
expansion of the universe is not
caused by a new physical field but is induced by a boundary condition,
which only manifests itself when the metric is expressed in terms of
Euclidian spacetime. While this leads to a $\Lambda$ parameter that is
constant across the 4D spacetime of the observable universe, its
magnitude depends on the age of the universe such that the value
of $\rho_\Lambda$ remains of the same order as the energy density of matter plus
radiation throughout cosmic history. 
The cosmic coincidence problem then
disappears. Without the use of any free 
parameters we derive a value of the cosmological constant 
parameter, $\Omega_\Lambda =67.2\,$\%, which
agrees within $2\sigma$ with the value obtained from observations
\citep{stenflo-planck2018arX}.

\section{Time Scale Introduced by the Cosmological
  Constant}\label{sec:timescale} 
The Einstein equation with cosmological constant $\Lambda$ can be
written in the form 
\begin{equation}\label{eq:einst}
R_{\mu\nu} -\Lambda\, g_{\mu\nu}={8\pi \,G\over c^4}
\,\bigl(\, T_{\mu\nu} -{1\over 2} \,g_{\mu\nu} T\,\bigr)\,. 
\end{equation}
Here we adopt the standard sign convention of \\ 
\citet[][]{stenflo-mtw1973}, with ($-++\,+$) for the
spacetime signature and a plus sign in front of the right-hand side. 

In the weak-field approximation  $R_{\mu\nu}\approx
-{1\over 2}\,\partial^2 g_{\mu\nu}\,$ if we adopt
the harmonic gauge. 
$\partial^2\equiv -(1/c^2)\,\partial^2/\partial t^2 +\nabla^2$ 
is the d'Alembertian 
operator $\Box^2$. Equation (\ref{eq:einst}) then describes
gravitational waves that propagate at the speed of light.  

In the Fourier domain the operator $\partial^2/\partial t^2$
corresponds to a frequency squared, $\omega^2$, which defines a
time scale $2\pi/\omega$. The combination
$(1/c^2)\,\partial^2 g_{\mu\nu}/\partial t^2-\Lambda\, g_{\mu\nu}$
for the left-hand side of 
Eq.~(\ref{eq:einst}) shows that $\Lambda$ corresponds to a frequency
scale $\omega_\Lambda$ and time scale $t_\Lambda$ defined by
\begin{equation}\label{eq:lamomtlam}
\Lambda\equiv \textstyle{1\over
  2}\,(\omega_\Lambda/c)^2\equiv 2\,[\pi/(c\,t_\Lambda)]^2\,.
\end{equation}

In standard cosmological treatments $\Lambda$ in Eq.~(\ref{eq:einst})
is generally moved to the right-hand side, where it 
formally appears as an
equivalent mass-energy density $\rho_\Lambda$, which in modeling is
represented by the dimensionless parameter $\Omega_\Lambda$. These
parameters are defined by 
\begin{equation}\label{eq:rholam}
\rho_\Lambda\equiv \Omega_\Lambda\rho_c\equiv {c^2\,\Lambda \over 8\pi \,G}\,, 
\end{equation}
where we have introduced the critical density 
\begin{equation}\label{eq:rhocrit}
\rho_c ={3\over 8\pi \,G \,t_H^2}\,,
\end{equation}
here expressed in terms of the Hubble time $t_H =1/H$, where $H$ is the
Hubble constant. $\rho_c$ represents the mean mass density in a
Friedmann universe with zero spatial curvature, which defines the
boundary between open and closed model universes. 

Combining Eqs.~(\ref{eq:lamomtlam})-(\ref{eq:rhocrit}) we obtain 
\begin{equation}\label{eq:tlamth}
{t_\Lambda\over t_H}={2\pi\over \sqrt{\,6\,\Omega_\Lambda}}\,.
\end{equation}
Inserting the observational value of $\Omega_\Lambda=0.685$ determined
by \citet{stenflo-planck2018arX}, we find $t_\Lambda /t_H \approx
3.10$ which, as will be shown below, implies a  value of $t_\Lambda $ that is very
close to the current conformal age 
$\eta_u$ of the universe. The distance $r_u=c\,\eta_u$ represents the radius of
the causal or particle horizon, and $\eta_u$ is the time that it would take
for a photon to travel this distance if the
universe would stop expanding. It is substantially longer than
the normal age of the universe, because the spatial points from which
light is emitted continually recede from us due to the cosmic
expansion. 

Instead of the temporal operator $\partial^2/\partial t^2$ we could
have used the same arguments for the corresponding spatial operators
and would have found that the observed value of $\Lambda$ corresponds to a
spatial scale $r_\Lambda$ that is very close to the magnitude of
$r_u$. While the discussion does not depend on the choice of temporal
or spatial coordinates, we choose to refer to time scales $t_\Lambda$ 
and $\eta_u$ for convenience. This should become clearer
when we develop the concepts in terms of the Robertson-Walker metric
in the next section. 

The circumstance that the observationally determined value of
$t_\Lambda$ nearly equals the conformal age $\eta_u$ would be an
astounding coincidence unless there is an
underlying physical reason for it. Since $t_\Lambda\sim 1/\sqrt{\Lambda}$ and
$\eta_u$ increases with the age of the universe, the coincidence would
only apply to our particular epoch and make us be highly privileged
observers, unless $\Lambda$ would scale with $1/\eta_u^2$ and thereby
conserve the apparent 
``coincidence'' for all cosmic epochs. In the following
two sections we will see how this is achieved.

\section{Boundary Condition in Euclidian Spacetime as the Origin of
  the Cosmological Constant}\label{sec:perbound}
Our aim here is to show that the cosmological constant has its origin
in a boundary condition that ties its value to the conformal age of
the universe. This boundary condition is 
hidden from view in the ordinary spacetime representation, but becomes exposed when
the treatment is extended to Euclidian spacetime. When applied to the event horizon of
black holes, the Euclidian approach offers a direct and elegant route to the derivation of the
expression for the temperature of Hawking radiation, as first shown by
\citet[][]{stenflo-gibbonsperryprl78}. While our objective is to 
apply it to the Robertson-Walker metric for the cosmological case, it
is helpful to begin with a look at the 
black hole case, to allow us to appreciate the underlying physics with
its differences and  similarities.

\subsection{Euclidian route to the Hawking
  temperature}\label{sec:hawking}
The Schwarzschild metric is given by 
\begin{equation}\label{eq:schmetric}
{\rm d}s^2 = -(1-r_s/r)\,c^2 {\rm d}t^2 +(1-r_s/r)^{-1} \,{\rm d}r^2+ \,r^2\,{\rm d}\Omega\,,
\end{equation}
where 
\begin{equation}\label{eq:schradius}
r_s ={2\,GM\over c^2}
\end{equation}
is the radius of the event horizon, and 
\begin{equation}\label{eq:domega}
{\rm d}\Omega={\rm d}\theta^2 +\sin^2\!\theta \,{\rm d}\phi^2
\end{equation}
is the surface element on the unit sphere. 

Let us now perform a Wick rotation in the complex plane to
transform ordinary time $t$ to Euclidian time $t_E= i\,t$, to make
time space like, thereby changing the signature of the metric from
($-++\,+$) to ($+++\,+$). In the immedate neighborhood of the event
horizon, where 
$\vert r-r_s\vert\ll r_s$, the metric can be written as 
\begin{equation}\label{eq:eucschmet}
{\rm d}s^2 = [(r-r_s)/r_s]\,c^2 {\rm d}t_E^2 +[(r-r_s)/r_s]^{-1}
\,{\rm d}r^2+ \,r_s^2\,{\rm d}\Omega\,.
\end{equation}
Following the treatment in \citet[][]{stenflo-bookzee2010}, we now
convert the metric near the horizon to the form 
\begin{equation}\label{eq:alphametric}
{\rm d}s^2 = R^2{\rm d}\alpha^2 \! +\,{\rm d}R^2+ \,r_s^2\,{\rm d}\Omega\,,
\end{equation}
where $r$ and $t_E$ have been replaced by two different coordinates, a 
distance parameter $R$ and an angular coordinate $\alpha$. The way to
obtain this form is to first define $R$ and a fiducial parameter
$\gamma$, which later will be disposed of, through $(r-r_s)/r_s\equiv
\gamma^2 R^2$. Since ${\rm d}r/{\rm d}R=2R\,r_s\gamma^2$, we find that
the second term on the right-hand side of Eq.~(\ref{eq:eucschmet})
is $[(r-r_s)/r_s]^{-1}\,{\rm d}r^2=(2\,r_s\gamma)^2\,{\rm d}R^2$,
which becomes ${\rm d}R^2$ as required by Eq.~(\ref{eq:alphametric})
provided that we choose $\gamma=1/(2\,r_s)$. This finally gives us the
desired form of Eq.~(\ref{eq:alphametric}) for the metric, if we
define ${\rm d}\alpha$ through 
\begin{equation}\label{eq:dalphars}
{\rm d}\alpha\equiv \gamma\, c\,{\rm d}t_E ={c^3\over 4\,GM}\, {\rm d}t_E\,.
\end{equation}

Equation (\ref{eq:alphametric}) describes a 4D space that has an 
$\vect{E}^2\otimes \vect{S}^2$ representation, 
i.e., it is a direct product between a Euclidian plane ($\vect{E}^2$,
represented by the first two terms in Eq.~(\ref{eq:alphametric}),
which describe the line element in polar coordinates in the Euclidian plane) and
a 2-sphere ($\vect{S}^2$), represented by the surface element ${\rm d}\Omega$ of the
unit sphere. 

Because $\alpha$ is an angular coordinate with period $2\pi$ in the
Euclidian plane, Euclidian time has a periodicity of
$\Delta t_E =2\pi/(c\,\gamma)$ according to
Eq.~(\ref{eq:dalphars}). The metric thus returns to the same state
after the $\Delta t_E$ interval. 
According to Euclidian field theory, which among other areas has important
applications in condensed matter physics \citep[cf.][]{stenflo-mccoy94},
the transition amplitude between two states of a 
quantum field in
Euclidian spacetime, when the initial and final states
are the same,  becomes the partition function of statistical
mechanics in ordinary spacetime. In particular this implies that
across the Euclidian time interval $\Delta t_E=i\,\Delta t$ the
oscillating phase factor $\exp(i\,\omega\, \Delta
t_E)=\exp(-\omega\,\Delta t)$ can be identified in ordinary spacetime with the
Boltzmann factor $\exp[-\hbar\, \omega/(k_B T)]$. 

This identification establishes a link between Euclidian field theory
and thermodynamics, allowing us to assign a temperature to the event
horizon of the black hole, as a consequence of the periodic boundary condition for the
metric in Euclidian spacetime. With Eq.~(\ref{eq:dalphars}) we find 
\begin{equation}\label{eq:bhtemp}
T={\hbar\over k_B\,\vert\Delta t\vert}={\gamma\,\hbar\, c\over 2\pi\,k_B}={\hbar
\,  c^3\over 8\pi\,GM\,k_B}\,.
\end{equation}
This agrees exactly with the expression for the Hawking temperature
that has been derived by other, independent methods.

\subsection{Euclidian route to the cosmological boundary
  condition}\label{sec:euclbc} 
The Robertson-Walker metric in the case of zero spatial curvature is
given by 
\begin{equation}\label{eq:robwalk}
{\rm d}s^2 = -c^2 {\rm d}t^2 +a(t)^2 \,({\rm d}r^2+ \,r^2\,{\rm d}\Omega)\,,
\end{equation}
where $a(t)$ is the scale factor and $r$ is the comoving distance. 
Let us now introduce Euclidian conformal time $\tau$ through the
definition 
\begin{equation}\label{eq:defeucltime}
{\rm d}\tau\equiv i\,c\,{\rm d}\eta\equiv  i\,c\,{\rm d}t\,/a\,,
\end{equation}
where $\eta$ is ordinary conformal time. Note that the speed of light
$c$ has been incorporated in this definition, so that $\tau$ has the
dimension of space. The metric now acquires the conformal Euclidian form 
\begin{equation}\label{eq:euclmetric}
{\rm d}s^2 = a(\tau)^2 \,( {\rm d}\tau^2 \! +\,{\rm d}r^2+ \,r^2\,{\rm d}\Omega)\,.
\end{equation}

According to Eq.~(\ref{eq:euclmetric}) all components of the conformal
metric are proportional to $a(\tau)^2$. Because monotonous,
e.g. exponential, behavior in ordinary time becomes circular
behavior in Euclidian time, we expect the Euclidian
spacetime metric to have periodic boundary conditions and develop
resonances, in particular since Euclidian time $\tau$ is bounded
between $\tau=0$ (Big Bang) and $\tau=\tau_u=i\,c\,\eta_u=i\,r_u$, where
$\eta_u$ is the conformal age of the observable universe and $r_u$ is
the radius of the causal (particle) horizon at the given
epoch. 

The existence of a metric resonance with respect to Euclidian time
implies that 
\begin{equation}\label{eq:a2sime}
a(\tau)^2\sim e^{i\,\omega\,\tau/c}\,,
\end{equation}
where $\omega$ is the resonant frequency that characterizes the
periodicity. Because Euclidian time is circular and bounded, and 
the circle should close by the available finite time line $\tau_u$ after an angle of
$2\pi$ for reasons of topological consistency (as clarified more below
in terms of the corresponding metric structure with polar angle $\alpha$ in
Eqs.~(\ref{eq:newalphkdr}) and (\ref{eq:robwalknewform})), the resonant
frequency needs to be 
\begin{equation}\label{eq:resonfreq}
\omega/c =2\pi/\tau_u\,.
\end{equation}

Note that the metric components are represented by the square of the
scale factor. Therefore the unsquared scale factor $a(\tau) $ has an 
oscillating phase factor with resonant frequency $\omega/2$. 

We will now introduce new coordinates $R$ and $\alpha$, with the aim
of converting our metric in Eq.~(\ref{eq:euclmetric}) to a form
resembling that of
Eq.~(\ref{eq:alphametric}) for the black hole case. With the
definition 
\begin{equation}\label{eq:requiva}
R\equiv a^2 
\end{equation}
and Eq.~(\ref{eq:a2sime}) we find 
\begin{equation}\label{eq:droadtau}
{\rm d}R= i\,{\omega\over c}\,R\,{\rm d}\tau\,.
\end{equation}
Next we define the angular coordinate $\alpha$ through 
\begin{equation}\label{eq:alphkdr}
{\rm d}\alpha \equiv i\,{\omega\over c}\,\,{\rm d}r\,.
\end{equation}
It may at first give the impression that we are dealing with imaginary angles, but
this is not the case as can be seen if we insert the expression
of Eq.~(\ref{eq:resonfreq}) for $\omega$ and recall that $\tau_u =
i\,r_u$. We then find that 
\begin{equation}\label{eq:newalphkdr}
{\rm d}\alpha =2\pi\,{\rm d}r/r_u =2\pi\,{\rm d}\eta/\eta_u\,.
\end{equation}

With these definitions the Robertson-Walker metric can be brought to the form 
\begin{equation}\label{eq:robwalknewform}
{\rm d}s^2 ={1\over R}\!\left({r_u\over 2\pi}\right)^{\!\!2}\!(\,R^2{\rm d}\alpha^2 \! +\,{\rm
  d}R^2\,+ \,\alpha^2R^2\,{\rm d}\Omega\,)\,.
\end{equation}
Like in the black hole case Eq.~(\ref{eq:robwalknewform}) describes a 4D space
with an $\vect{E}^2\otimes \vect{S}^2$ representation, a direct
product between a Euclidian plane and a 2-sphere (which now includes
an additional conformal factor). It obeys a
periodic boundary condition, because the angular coordinate $\alpha$
returns the metric to its initial state after an angular interval of
$2\pi$, which according to Eq.~(\ref{eq:newalphkdr}) corresponds to a
conformal time interval $\eta_u$. 

If the frequency $\omega$ were smaller than the value given by
Eq.~(\ref{eq:resonfreq}), then all angles in the Euclidian
$\vect{E}^2$ plane would not be covered by the available time line,
and if instead $\omega$ were larger, parts of the plane would be
multiply covered. Only the choice of Eq.~(\ref{eq:resonfreq}) ensures
unique, exact coverage. 

Note that while the metric returns to its original state after one
revolution of $2\pi$, the phase factor for the unsquared scale factor
$a$ requires two revolutions. In this respect it behaves like a spinor
in Euclidian spacetime. 

In Sect.~\ref{sec:pbclam} we will show that $\omega$, which represents
a resonance in Euclidian spacetime,  
generates a cosmological constant $\Lambda$ that is
proportional to $1/\eta_u^2$, when we convert back to ordinary
time. The expression for $\Lambda$ is identical 
to that of Eq.~(\ref{eq:lamomtlam}) if we identify the time interval
$t_\Lambda$ with the conformal age $\eta_u $ of the universe. This raises a
non-trivial conceptual issue that will be addressed in the next
subsection. How can $\Lambda$ remain constant 
across all of the observable universe at a given epoch while being a
function of the age of the universe\,?

\subsection{Global constraint versus physical
  field}\label{sec:globconst}
In distance -- redshift space observers are by definition always
located at redshift $z=0$, although their location in ordinary space may be 
entirely arbitrary. The choice of observer always defines the zero point of the
redshift scale and the age of the universe. All local physics that any
observer can ever deal with takes place at $z=0$. In contrast,
information about non-local objects (with $z>0$) can only be  
inferred from their measured spectral properties via a cosmological
model. 

The model used for the inference is based on Einstein's
theory of gravity, which describes the relation between the physical
fields, represented by the energy-momentum tensor $T_{\mu\nu}$, and the
geometry of spacetime, which is expressed as a function $G_{\mu\nu}$
of the metric $g_{\mu\nu}$. The functional form of $G_{\mu\nu}$ is
governed by the requirement that $T_{\mu\nu}$ of
the physical fields must be conserved. A cosmological constant term
$\Lambda g_{\mu\nu}$ can be added to the Einstein equation without
being affected by the conservation constraint, because it is divergence free as
long as $\Lambda$ is a true constant. It is therefore absent from the
energy-momentum conservation equation, which determines the properties
of the physical fields. 

In our present theory $\Lambda$ is not a physical field but expresses
a global cyclic property or resonance of the metric. As a consequence the magnitude of
$\Lambda$ is tied to the finite age $\eta_u$ of the universe. While $\Lambda$ is
a function of $\eta_u$, it does not depend on look-back
time, which is a quantity that is inferred via a cosmological
model. Because the constant parameter $\Lambda$ is an ingredient of
the model used, look-back time depends on $\Lambda$ but does not
influence its value. 

Our metric resonance has some superficial similarity to the Casimir
effect, according to which the energy density of the quantum vacuum is changed as a
result of boundary conditions for the vacuum electromagnetic modes
inside a volume bounded by electrically conducting parallel plates. The
change of the vacuum energy occurs 
because the boundary constraints only allow the existence of a
discrete set of modes. The effect is not a function of position along the
dimension perpendicular to the plates, it applies globally to the
space between the plates. Similarly, in our metric
resonance case, the resonant frequency induced by the boundary
condition is constant and representative of the whole 4D cavity of the
observable universe at epoch
$\eta_u$.  When in the Casimir case the plate separation is changed, the
vacuum energy within the entire volume between the plates changes 
without generating spatial gradients. Similarly, in our metric case
when the value of $\eta_u$ changes (the universe grows older), the
resonant frequency (which governs the magnitude of $\Lambda$) changes, but
its value is independent of 4D position within the observable universe
at epoch $\eta_u$. 

Instead of the Casimir effect we could also use a violin string as an
example. The discrete frequencies that it generates because of the
boundary conditions are not localized along the string but represent a
global property of the string. If we would stretch the
string, the frequencies would change but only be functions of the
string length, without spatial gradients. 

In spite of this analogy, the
origin of $\Lambda$ is fundamentally different. It
is not an expression of some form of cosmic Casimir effect, in
particular because the
nature of the boundary conditions is different. The
Casimir effect and the violin string are subject to Dirichlet boundary
conditions, which constrain the oscillating amplitudes to vanish at the
boundaries. When the end points are clamped down, the string length
corresponds to half a wavelength for the fundamental mode, an angular
interval of $\pi$. In contrast the metric has a resonance because
Euclidian conformal time with a temporal string of length of $\eta_u$ obeys
periodic boundary conditions with  
fundamental angular period $2\pi$. There is no physical justification
for applying Dirichlet boundary conditions 
to $\eta_u$. Such boundary conditions can also be ruled out on the
grounds that a temporal string length of $\pi$ (half a wave length) would lead to
a $\Lambda$ in discord with the observed value. The observational
constraints require a resonant frequency 
that corresponds to a string length of $2\pi$, not $\pi$.

\subsection{Classical and quantum aspects}\label{sec:classquant}
Our explanation of the origin of the cosmological constant has so far
only made use of classical physics, at least in the 
sense that Planck's constant $\hbar$ does not appear in the expression for
$\Lambda$. When using Euclidian spacetime to derive the expression
for the Hawking temperature of black holes in Sect.~\ref{sec:hawking}, 
$\hbar$ appeared because of the identification made between the
oscillating factor in Euclidian spacetime and the Boltzmann factor in
ordinary spacetime. Similarly, if $\omega_u$ is the 
resonant frequency of Eq.~(\ref{eq:resonfreq}) that is associated with
the finite Euclidian conformal time
string $\tau_u=i\,c\,\eta_u$, then we can make the identification of
$\exp(i\,\omega_u\,\tau_u/c)=\exp(-\omega_u\,\eta_u)$ with
$\exp[-\hbar\,\omega_u/(k_B T_u)]$ to obtain the 
temperature $T_u$ that is associated with 
$\eta_u$. It gives us $T_u=\hbar/(k_B \eta_u)$, which has the stupendously small value of 
about $10^{-29}$\,K due to the gigantic value of $\eta_u$. This
expression is the same as the corresponding expression of
Eq.~(\ref{eq:bhtemp}) for the black hole temperature if we replace $\Delta t$ with $\eta_u$. 

Equation (\ref{eq:bhtemp}) and $T_u=\hbar/(k_B \eta_u)$ can in fact be
seen as direct consequences of Heisenberg's uncertainty
principle $\Delta E\,\Delta t \approx \hbar/2$, if we identify $\Delta
E$ with $k_B T_u/2$ and $\Delta t$ with $\eta_u$. The inverse scaling
between $T_u$ and $\eta_u$ allows us to easily scale $T_u$ back to the Planck
era. In units of the Planck time $5.4\times 10^{-44}\,$s the current
conformal age of the universe is approximately $10^{61}$. This implies
that $T_u$ in the Planck era was approximately 
$10^{-29}\times 10^{61}=10^{32}\,$K, in agreement with the value of $m_P\,
c^2/k_B$ that is generally defined as the Planck temperature, where
$m_P=(\hbar\, c/G)^{1/2}\approx 22\,\mu$g is the Planck mass.

\subsection{From periodic boundary condition to $\Lambda$}\label{sec:pbclam} 
Although the temperature concept or
Planck's constant do not enter in the expression for $\Lambda$, 
these concepts are needed to estimate the amplitude of the resonance
in order to show
that the weak-field approximation can safely be used except in the
immediate neighborhood of the Planck era. Within classical physics we have
no principle that can be used to constrain the amplitude. In the
previous subsection we described two ways in which quantum physics
determines the mode amplitude: (1) via the relation between
Euclidian field theory and statistical mechanics, and (2) through
application of the Heisenberg uncertainty principle to the finite time
string $\eta_u$. The two ways of dealing with this issue lead to the same result. 

The temperature of $T_u\approx 10^{-29}\,$K that was shown to be
associated with our resonant frequency $\omega_u$ is insignificant in
comparison with the CMB temperature and completely
irrelevant to the present evolution of the universe. The related mode energy
$\hbar\,\omega_u$, which like $T_u$ scales with $1/\eta_u$, has a present
value that is $10^{-61}$ when expressed in
units of the Planck energy. While insignificant at present, this
incredibly tiny number increases as 
we go back in time to become unity in the Planck era. It represents the
relative amplitude by which the metric fluctuates and can be
interpreted in the Newtonian limit as a potential energy. As long as this
amplitude is $\ll 1$, we are in the regime where the weak-field approximation
for the Einstein equation is valid. This weak-field criterion is
satisfied for all times except for the non-linear regime in the
vicinity of the Planck era when 
the universe is younger than about $10^{-41}\,$s (at which time the
resonant amplitude is about 0.005). 

Because the Euclidian spacetime metric obeys a periodic boundary
condition with a resonant frequency $\omega_u
=2\pi\,c/\vert\tau_u\vert=2\pi/\eta_u$, it 
acquires the behavior of a harmonic oscillator with equation $\partial^2
g_{\mu\nu}/\partial \tau^2+(\omega_u/c)^2\,g_{\mu\nu}=$ source terms (the physical
fields that are represented by $T_{\mu\nu}$). We recall that the $\tau$ coordinate is
the conformal Euclidian time that was defined in Eq.~(\ref{eq:defeucltime}). 

At first glance it may appear strange that the metric could have
a circular behavior, governed by a phase factor
$\exp(i\,\omega_u\tau/c)$ that returns the metric to its
initial value after a certain interval. It does not at all
seem to resemble the universe that we live in. This confusion gets
resolved if we recall that the circular behavior refers to Euclidian
time, which is not the time that we experience. When we convert back
to ordinary time, the oscillating phase factor becomes a de Sitter
factor that describes an exponential expansion of the universe,
exactly the behavior that we expect $\Lambda$ to represent. Both 
an exponentially decaying and expanding behavior are formally allowed,
but with 
the initial condition that the universe has evolved from a dense state
(with small scale factor), the decaying
solution can be excluded. An empty universe 
would experience an exponential de Sitter expansion, but because
physical fields also exist, the $\Lambda$ contribution is just
one of the terms that govern the behavior. 

When converting back from Euclidian conformal time $\tau$ to ordinary
conformal time $\eta$, the relative signs of the two terms in the oscillator
equation changes (because of the squaring of imaginary $i$), and we get $\partial^2
g_{\mu\nu}/\partial \eta^2-\omega_u^2\,g_{\mu\nu}=$ source
terms. Because of the sign change this is no longer an oscillator
equation with a periodic solution. 

To relate $\omega_u^2$ to $\Lambda$ we recall from Sect.~\ref
{sec:timescale} that in the weak-field approximation the left-hand
side of the Einstein equation becomes $R_{\mu\nu} -\Lambda\,
g_{\mu\nu}\approx [1/ (2c^2)]\,\partial^2 g_{\mu\nu}/\partial t^2 -\Lambda\,
g_{\mu\nu}$, if we would ignore the expansion scale factor $a(t)$, and let $t$
represent ordinary time. This expression preserves its form in an
expanding universe with scale factor $a(t)$ if we instead of time $t$
use conformal time $\eta$ and redefine the metric so that
$g_{\mu\nu}\equiv a(\eta)^2\, g_{\mu\nu,\,{\rm non\,exp}}$, where
$g_{\mu\nu,\,{\rm non\,exp}}$ refers to the metric in the
non-expanding case (with $a(\eta)\equiv 1$). Multiplying all terms of the
equation by $2c^2$, the 
left-hand side becomes $\partial^2 g_{\mu\nu}/\partial \eta^2
-2c^2\Lambda\,g_{\mu\nu}$. Comparing with our corresponding expression
that was derived from the metric resonance $\omega_u$, we find that 
\begin{equation}\label{eq:lamomega}
\Lambda =\textstyle{1\over
  2}\,(\omega_u/c)^2 = 2\,[\pi/(c\,\eta_u)]^2\,,
\end{equation}
which is identical to Eq.~(\ref{eq:lamomtlam}) if we replace $t_\Lambda$ by $\eta_u$.

\section{Unique Solution for the Value of $\Omega_\Lambda$}\label{sec:unisol}
The most convenient way to parametrize $\Lambda$ for modeling purposes
is in terms of the dimensionless parameter $\Omega_\Lambda$, which was
defined in Eq.~(\ref{eq:rholam}). It 
represents the fraction of the critical density
$\rho_c$ that is in the form of $\rho_\Lambda$, the vacuum energy density
version of the cosmological constant. Using
Eqs.~(\ref{eq:rholam}) and (\ref{eq:lamomega}), we get 
\begin{equation}\label{eq:omlamtheo}
\Omega_\Lambda\,=\,{2\over 3}\left[{\pi\over x_u(\Omega_\Lambda)}\right]^2\,.
\end{equation}
The dimensionless function $x_u$ represents the conformal age of
the universe in units of the Hubble time: 
\begin{equation}\label{eq:xcdef}
x_u\equiv \eta_u/t_H\,.
\end{equation}
Because it is a function of $\Omega_\Lambda$, as indicated
explicitly in Eq.~(\ref{eq:omlamtheo}) and as will be brought out in the
expressions below, $\Omega_\Lambda $ gets uniquely defined by
Eq.~(\ref{eq:omlamtheo}). 

From the definition of conformal time we obtain  
\begin{equation}\label{eq:xcinteg}
x_u ={1\over t_H}\int_0^{t_u}{{\rm d}t\over a}=\int_0^\infty {{\rm d}z\over E(z)}\,,
\end{equation}
where $t_u$ is the age of the universe in ordinary time units and $z$ is
the redshift, while 
\begin{equation}\label{eq:ez}
E(z)=[\,\Omega_M\,(1+z)^3 \,+\,\Omega_R\,(1+z)^4\,+\,\Omega_\Lambda\,]^{1/2}
\end{equation}
and 
\begin{equation}\label{eq:omsum}
\Omega_M\,= 1\,-\,(\,\Omega_R +\Omega_\Lambda) 
\end{equation}
with our assumption of zero spatial curvature. $\Omega_M$ represents
the matter density (including dark matter) in units of $\rho_c$. The
corresponding density parameter for radiation is 
\begin{equation}\label{eq:omegar}
\Omega_R\,={u_R\over \rho_c\,c^2}\,.
\end{equation}
The radiation energy density $u_R$ is the sum of the contributions
from the photon and neutrino backgrounds: 
\begin{equation}
u_R =a_T\,T^4\left[1+{7\over 8}\left({4\over
 11}\right)^{\!\!4/3}\!\!N_\nu\right]\,=1.681\, a_T\,T^4
\label{eq:ur}\end{equation}
\citep[cf.][]{stenflo-bookpeebles1993}. The number of neutrino families
$N_\nu =3$, $T=2.725\,$K is the measured temperature of the cosmic
microwave background, and $a_T$ is Stefan's constant. 

The above set of equations are sufficient to allow us to solve
Eq.~(\ref{eq:omlamtheo}), which gives us a unique value for
$\Omega_\Lambda$, namely 67.2\,\%. This is within $2\sigma$ of the
most recent value of $68.5\pm 0.7$\,\%\ that has been 
derived from observations \citep{stenflo-planck2018arX}. 

The dependence of the solution on the current value of the CMB
temperature is small. If we would neglect the radiation
contribution $\Omega_R$ altogether we would get
$\Omega_\Lambda =66.3\,$\%, which only differs by $3\sigma$ from the
observed value. The situation is however quite different if we instead
would disregard the matter contribution $\Omega_M$ as we do for the 
radiation-dominated era of the early universe. In this case
$\Omega_\Lambda$ is as high as 93.1\,\%. 

In standard cosmology $\Omega_M$, $\Omega_\Lambda$, and the spatial
curvature can be neglected in the early universe. The scale factor
$a(t)$ is then governed by the Friedmann solution for a radiation-dominated
universe, for which the ratio $t_u/t_H$ between the age of the
universe and the Hubble time is 0.5. The cosmology that follows from
our solution of Eq.~(\ref{eq:omlamtheo}) is however quite different
for the early universe,
since $\Omega_\Lambda=93.1\,$\%\ and therefore highly
significant. Calculating $t_u/t_H$ from 
\begin{equation}\label{eq:tuthinteg}
{t_u\over t_H} =\int_0^\infty {{\rm d}z\over (1+z)\,E(z)}
\end{equation}
gives us 1.044 in this case, which implies an expansion rate in the early
universe that is faster than that of the standard Friedmann case by the
factor $1.044/0.5\approx 2.09$. 
The faster expansion rate has implications for our interpretation and
modelling of the physical processes in the early universe. This has 
consequences for the values of the model parameters that are needed to
fit the observables, like the baryon density parameter $\Omega_B$
needed for the BBN (Big Bang nucleosynthesis) predictions to match the
observed abundances of the light elements, or for the array of other
parameters needed to fit the observed CMB spectrum. 

The presence of a cosmological constant implies the existence of both
a time scale and a spatial scale. While we have identified the time scale as
the conformal age $\eta_u$ of the universe, the
corresponding spatial scale is $r_u =c\,\eta_u$, which represents the
radius of the causal or particle horizon at the given epoch. Equation
(\ref{eq:lamomega}) shows how $\Lambda$ can be represented directly in
terms of this spatial scale: $\Lambda =2\,(\pi/r_u)^2\,$. The
magnitude of $\Lambda$ thus always tracks the size of the horizon. 
It is just a matter of convenience whether we choose to refer to this size in
terms of  temporal or spatial units (as long as we use conformal
coordinates). 

The tracking $\Lambda$ term will have no significant direct 
influence on structure formation provided that the considered spatial and
temporal scales are much smaller than the horizon scale. For this
reason the effect on galaxy formation should be minor. However, the formation
of the large-scale structures and all other physical processes
take place in an expanding spacetime arena that is
governed by the evolving scale factor $a(t)$, which is different from
that of standard cosmology. To avoid confusion with the non-local concept of
look-back time we need to remember that the time $t$ in $a(t)$ is the
local, dynamical time scale that is experienced by a comoving
observer, and which represents the age of the universe for that
observer. This means that $t\equiv t_u$, or that $\eta\equiv \eta_u$ 
if we use conformal coordinates. The expansion rate, ${\dot
  a}(t)$, is driven by three sources: the energy
densities of matter and radiation, and the $\Lambda$ term. Because
$\Lambda$ represents a non-local effect that has its origin in a global
constraint and is therefore expressed by a global integral,
Eq.~(\ref{eq:xcinteg}), which is
used in Eq.~(\ref{eq:omlamtheo}), we have to solve an
integro-differential equation to obtain the solution for $a(t)$. This
solution provides the framework for the applications of the theory, in
particular for the tests when confronting it with observational
data.

\section{Conclusions}\label{sec:concl}
As the value of the cosmological constant is tied to the age of the
universe in our theory, there is no cosmic coincidence problem, no
violation of the Copernican principle. 
The contribution from the $\Lambda$ term to the cosmic expansion rate
has always been of the same order of magnitude as the contribution
from the physical fields (matter plus radiation), 
and it will remain so in the future. Therefore
we are not privileged observers. 

The periodic boundary condition for the metric, which is the origin of
the $\Lambda$ term, only manifests itself within the Euclidian
spacetime representation. The expression for the Hawking temperature
that is obtained with the Euclidian spacetime approach 
is identical to the corresponding expression obtained by
entirely independent methods, e.g. by \citet[][]{stenflo-hawking74} in
his discovery paper. This agreement supports the validity of the
approach, in spite of the absence of observational verifications of
the Hawking temperature. In our cosmological case the Euclidian
spacetime technique gets additional validation by predicting a value
for $\Lambda$ that closely agrees with the observed value. 

The circular, periodic property of the metric in Euclidian spacetime
becomes an exponentially evolving property in ordinary spacetime, a de
Sitter evolution, which is tempered by the comparable contributions from the
physical fields in a way that preserves the zero
spatial curvature of the large-scale metric. 

Without the use of any free fitting parameters we have
derived a value of $\Omega_\Lambda =67.2\,$\%\ that is within $2\sigma$ of the
observed value, an agreeement that can hardly be dismissed as a mere
``coincidence''. The uniqueness of our solution for $\Omega_\Lambda$
without the use of free parameters implies that there does not exist any parallel
universes with other values of the cosmological constant. Nature did
not have a choice, because logical consistency excludes alternative
possibilities. 

Our theory implies an expansion history, governed by the scale
  factor $a(t)$, which is significantly different from that of standard
cosmology. The different expansion history provides the framework to be used
when the theory is to be tested by comparing its predictions 
with the various observational constraints. 
 
As the $\Lambda$ term has remained significant
throughout the earlier history of the universe and therefore has led
to a different expansion rate, there will be
implications for the interpretation of events that took place in the
early universe, like structure formation, the processes that
  generated the CMB spectrum, or the BBN (Big
Bang nucleosynthesis). One such implication was explored in
\citet{stenflo-s2019a}, where it was shown that the different expansion
rate in the present theory would need to be compensated for by significantly
raising the baryon density parameter $\Omega_B$ in order to preserve
agreement between the BBN predictions and the observed deuterium
abundance, which in turn may have consequences for our understanding
of the nature of dark matter. Similarly other cosmological parameters will need
adjustments because of the changed expansion rate in order to maintain
agreement with the powerful CMB constraints. These are some of the
issues that will need to be addressed in future work.


\bibliographystyle{spr-mp-nameyear-cnd}  

\end{document}